# Blessing or Curse of Democracy?: Current Evidence from the Covid-19 Pandemic

**Authors:** Ryan P. Badman[1*], Yunxin Wu[2], Keigo Inukai[3], Rei Akaishi[1]
*Affiliations:*
[1]Center for Brain Science, RIKEN, Saitama, 351-0106, Japan
[2]Independent freelance researcher, Boston, MA, USA
[3]Department of Economics, Meiji Gakuin University, Tokyo, 108-8636, Japan
[*]Corresponding author: ryan.badman113 (at) gmail.com

**ABSTRACT:**

**Background:** A major question in Covid-19 research is whether democracies handled the Covid-19 pandemic crisis better or worse than authoritarian countries. However, it is important to consider the issues of democracy versus authoritarianism, and state fragility, when examining official Covid-19 death counts in research, because these factors can influence the accurate reporting of pandemic deaths by governments. In contrast, excess deaths are less prone to variability in differences in definitions of Covid-19 deaths and testing capacities across countries. Here we use excess pandemic deaths to explore potential relationships between political systems and public health outcomes.

**Methods:** We address these issues by comparing the official government Covid-19 death counts in a well-established John Hopkins database to the generally more reliable excess mortality measure of Covid-19 deaths, taken from the recently released World Mortality Dataset. We put the comparison in the context of the political and fragile state dimensions.

**Findings:** We find (1) significant potential underreporting of Covid-19 deaths by authoritarian governments and governments with high state fragility and (2) substantial geographic variation among countries and regions with regard to standard democracy indices. Additionally, we find that more authoritarian governments are (weakly) associated with more excess deaths during the pandemic than democratic governments.

**Interpretations:** The inhibition and censorship of information flows, inherent to authoritarian states, likely results in major inaccuracies in pandemic statistics that confound global public health analyses. Thus, both excess pandemic deaths and official Covid-19 death counts should be examined in studies using death as an outcome variable.

**Funding:** R.P.B. and R.A. were funded by a research grant from the Toyota Motor Corporation (LP-3009219). K.I. was funded by JSPS KAKENHI Grant Number 19K21701 and 17H04780.

## MAIN TEXT

*"…there is a real risk that political scientists and economists will publish analyses that try to attribute morbidity and mortality to policy and politics without understanding the serious and highly political limitations on data about COVID-19 infections and attributable mortality."*

-Greer et al. (*1*)

## Introduction

The Covid-19 pandemic (2019-2021) is an ongoing global public health catastrophe, with surprisingly disparate outcomes across countries (*2*, *3*). Despite many countries having comparable medical infrastructure and technology, some appear to contain the virus effectively while others have had the pandemic spiral out of control (*4–11*). Naturally, this has broadly led researchers to conclude that social variables such as culture, social norms, communication strategies, the type of government, prior pandemic experience, etc. must play a crucial role in Covid-19 pandemic outcomes in addition to medical infrastructure and technology (*1*, *12*, *13*). Historical public health research beyond the Covid-19 pandemic supports the view that social variables are quite important in pandemic management, and this general direction of research is a worthy endeavor (*14–16*).

However, there is a risk in such social science analyses, especially at the global scale, of focusing on too few relevant sociopolitical variables and then reaching erroneous conclusions (*1*). It is broadly recognized that pandemic outcomes are influenced by a complicated range of socioeconomic, institutional, historical, geographic, and cultural factors, confounding correlation analyses as well as causal analyses (*12*, *17–25*). Furthermore, especially when looking across countries with a large range of government types and technological development stages, official public health records based on government reports may be influenced by multiple factors (*10*, *26*) or directly tampered with by unethical governments to intentionally or unintentionally improve their own self-image (*27*).

Towards this important research direction, Narita et al. recently published the pre-print (*28*) (not yet peer-reviewed to our knowledge), "Curse of Democracy: Evidence from 2020" (arXiv:2104.07617) using official Covid-19 death numbers from national governments, which are collected and made available through the Covid-19 Data Repository by the Center for Systems Science and Engineering (CSSE) at Johns Hopkins University (*2*). In their paper, Narita et al. claimed that democracy 'caused' more Covid-19 deaths than authoritarian governments during the first year of the pandemic: there are higher rates of reported deaths in democratic countries than in authoritarian countries. This work follows earlier papers that reached similar conclusions favoring the pandemic responses of authoritarian countries during the first wave of the Covid-19 pandemic in early 2020 (*29–31*). However, there is a potential problem in the analyses presented in these studies. Authoritarian countries are frequently suspected to be misrepresenting their Covid-19 death tolls (*3*, *27*, *32*). Thus, using official government-reported death counts as an outcome variable to study the effects of government orientation is highly problematic for any Covid-19 study pursuing this research direction. In fact, all authoritarian regimes have general tendencies of information flow inhibition and censorship as hallmark features, which can prevent the accurate reporting of COVID-19 deaths. In contrast, not all authoritarian regimes have an effective coordinated public health authority (*1*, *33*), which can help both in more accurate reporting of COVID-19 data as well as in improving the containment of pandemic spread.

For the indices of democracy across countries, which are used to measure the degree of authoritarian versus democracy tendencies, we use the metrics created by the Economist Intelligence Unit (EIU) (2021) (*34*) and the Center for Systemic Peace (Polity5 Project) (for most countries the last update was 2018) (*35*). We also employ the Center for Systemic Peace's "state fragility index" (*35*, *36*). The state fragility index is a broader score of legitimacy and effectiveness in the four categories of security, political, economic, and social (*36*), and may provide a more comprehensive picture of how well a government can provide a stable society beyond the democracy-authoritarian dimension alone (this state fragility index is often used in considerations of foreign aid and investment) (*36–39*). Furthermore, for a better estimate of the real death rates due to the COVID-19 infections, we instead analyze national excess deaths during the pandemic. Excess deaths during the pandemic period are increasingly accepted to be a more reliable general measure of Covid-19 deaths over official government death counts (*3*, *27*, *40–44*). This movement towards excess deaths for Covid-19 mortality research is motivated by a variety of factors. For example, countries vary tremendously in their definitions of what constitutes an official Covid-19 "death", as well as in their capacity to test for Covid-19 (*26*). In contrast, excess deaths can be used to statistically determine an anomalous increase in deaths within a month or a year, by subtracting the baseline death rates from prior years (assuming no major war, and roughly consistent death rates from more routine diseases, violent crime, etc.) (*43*, *45*).

To obtain excess deaths counts, we employ the most up-to-date and extensive database for excess mortality, the new World Mortality Dataset (*3*), which is used by the Economist to report excess deaths for example (*46*), and is referenced within several recent research studies (*28*, *47–52*). Approximately 90 countries are available within this database, but many countries (including India and China) still lack the institutional infrastructure to track excess deaths, according to government communications reported by Karlinsky & Kobak (*3*). Our goal of this article is not to argue what the exact death toll is in various countries, but to show that official Covid-19 death counts as reported by national governments may grossly underestimate Covid-19 death counts, particularly in authoritarian and/or fragile state governments.

We will use the new World Mortality Dataset to demonstrate that official government death counts are biased by political systems. Indeed, as presented below, we find that the World Mortality Dataset (*3*) provides clear evidence that countries with lower democracy indices and a higher fragile state index are substantially more likely to undercount the number of Covid-19 deaths.

## Methods

We briefly describe our analysis of the comparison between the World Mortality Dataset of excess death counts during the Covid-19 pandemic (*3*) and the John Hopkins CSSE database of official Covid-19 deaths as reported by national governments (*2*). In our main analysis, we first calculate the ratio of undercounted deaths ("undercount") by dividing the number of Covid-19 pandemic excess deaths (starting in January 2020) from the World Mortality Dataset (*3*), by the number of official government-reported Covid-19 deaths per country from the John Hopkins CSSE database (*2*), as summarized by the following simple equation:

$$Undercount = \frac{Excess\ Death\ Count}{Official\ Death\ Count}$$

Undercount values here are actually *underestimates*. The official death counts presented in the John Hopkins CSSE database are generally a few weeks/months ahead of the World Mortality Dataset excess death counts. Then, the death value in the numerator of the undercount ratio should actually be larger due to this data lag. For example, the John Hopkins CSSE database with official government numbers is updated daily with official Covid-19 counts from every country, and it is easy to query this database to give all the official counts in the world for any date. On the other hand, the more recently released (spring 2021) World Mortality Dataset of cumulative Covid-19 pandemic excess death does not have the feature where you can easily look at all excess deaths on any date, as many countries do not release excess mortality counts daily (*3*). Currently, instead the World Mortality Dataset primarily presents the most current update for each country's (or region's) estimate of the total number of excess deaths they experienced during the pandemic. Almost all countries gave their last update between December 2020 and early April 2021, a few slightly earlier than December 2020 (see Supplementary Information). Additionally, even in highly developed countries there is often country-dependent reporting lag that could be weeks to perhaps months in official national Covid-19 death counts (as well as in excess deaths reports) (*26*, *53*–*56*), which complicates any mortality study. Especially for global studies, where there are huge variations in medical record keeping and communication infrastructure across the world (as well in Covid-19 testing abilities), it is incredibly difficult to estimate the lag in either Covid-19 deaths or excess death reporting (*26*). Thus, as mentioned previously, by using official death counts from the latest possible time point in the excess death calculation interval, we hope to only risk underestimating the severity of the undercount value (rather than overestimating), to help avoid overstating any conclusions presented here.

For countries with low (~1000 or less) Covid-19 death counts and a negative (decrease) in excess deaths during the pandemic period relative to prior years, undercount values were set to “1” for this estimate due to there being no discrepancy in death reporting and to avoid negative inputs to logarithms. The entries set to an undercount of “1” were Australia, Denmark, Finland, Iceland, Jamaica, Japan, Malaysia, Mauritius, Mongolia, New Zealand, Philippines, Singapore, South Korea, Taiwan, and Uruguay. Again, note the goal of this work is not to argue the exact numbers of Covid-19 deaths in any given country but rather to estimate the magnitude of inaccuracies in the John Hopkins CSSE Covid-19 death counts.

The values of the World Mortality Dataset were checked with values of the more established Human Mortality Database (*57*) to confirm agreement between excess death values from the two datasets. We examined the data for the 30 overlapping countries between these two datasets. However, the Human Mortality Database primarily documents Western democracies, so could not be used in the full analysis (*3*). The values of the World Mortality Dataset were also checked against the very recently published University of Washington Institute for Health Metrics and Evaluation (IHME) dataset of estimated true Covid-19 related mortality per country(*58*). Significant though systematic differences exist between this IHME dataset, and the World Mortality Dataset (as well as other preliminary excess death datasets (*46*, *59*)), with consistently 1.5-2 times higher excess death estimates in IHME per country than the World Mortality Dataset. A notable concerning discrepancy between datasets we found is Japan, where the IMHE estimates the true Japan Covid-19 death count is over 10 times higher than reported by official Covid-19 death counts (*2*) and up to 5 times higher than other previously calculated excess deaths ranges for Japan across all prefectures (*44*, *60*). We suspect that the IMHE dataset has insufficiently corrected for the unusually elderly-heavy age structure in at least East Asian democracies (a frequent and well-established issue in Covid-19 death analyses) (*59*, *61*), and thus have not used this dataset in this work.

## Results

*Undercounting with Authoritarian and Fragile State Governments*

The undercount values are plotted against values of the indices of democracy and state fragility across countries (Fig. 1). Countries with lower democracy indices and higher state fragility indices are found to substantially undercount their Covid-19 deaths. Specific examples of country-level underreporting among countries with lower democracy indices are given in Table 1 as well.

Pearson coefficients for the undercount ratio versus each index are given in Fig. 1. The correlational analysis was also repeated without the well-performing 15 negative excess death countries mentioned in the Methods section (i.e. those manually set to log(1) = 0 in Fig. 1), finding Pearson correlation coefficients of -0.623 ($p \sim 10^{-8}$), -0.506 ($p \sim 10^{-5}$), and 0.550 ($p \sim 10^{-6}$) for the EIU democracy index, Polity5 democracy index, and state fragility index respectively, for the remaining countries.

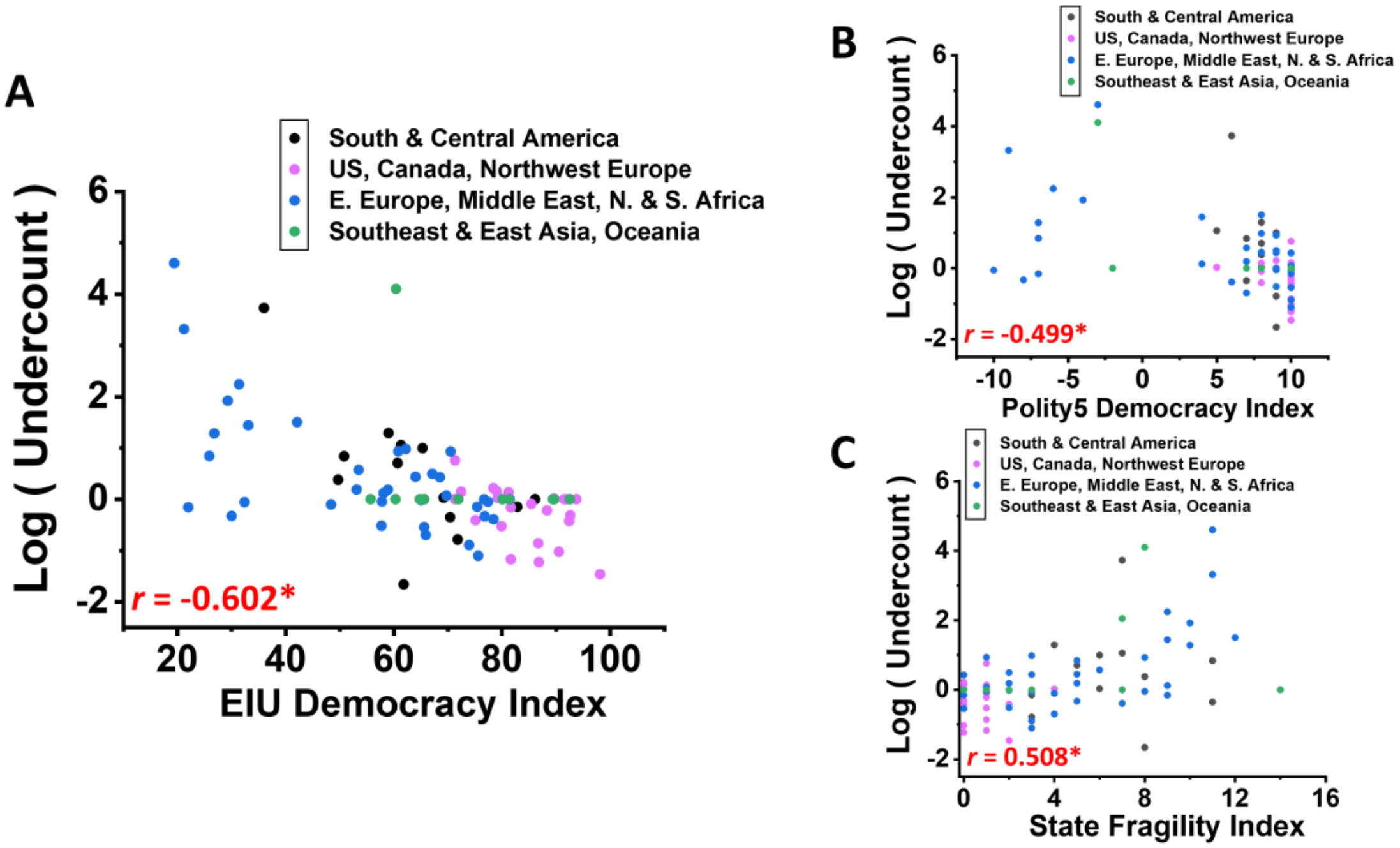

**Figure 1: The degree of governments undercounting deaths versus national indices of democracy and state fragility** The ratio for undercounted deaths ("undercount") is calculated by dividing the number of excess deaths from the World Mortality Dataset (*3*) by the number of official government reported deaths per country from the CSSE John Hopkins database (*2*). (A) The natural log of undercount ratio versus the EIU democracy index, with a strong negative Pearson correlation coefficient ($r$) observed ($p \sim 10^{-9}$) ($N$=83). (B) The natural log of undercount ratio versus the Polity5 democracy index, with a strong negative correlation observed ($p \sim 10^{-6}$) ($N$=80). In (A) and (B) a higher democracy index means a more democratic government, and a lower index means a more authoritarian

government (*34*). (C) The natural log of undercount ratio versus the state fragility index, with a strong positive correlation observed ($p \sim 10^{-6}$) ($N$=81). A higher state fragility index means a more unstable country.

| **Low Democracy Index Country** | **Official Government Deaths*** | **Estimated Excess Deaths During Pandemic (+/- SD)** | **Undercount Ratio** |
|---|---|---|---|
| Russia | 104,937 | 443,695 (+/-  29,020) ** | 4.23 |
| China-Hubei | 4,512 | 6,000-36,000 (+/-  ?)*** | ~1-8 |
| China-national | 4,636 | ? (+/-  ?) **** | ? |
| Azerbaijan | 4,235 | 15,309 (+/-  1,222) ** | 3.61 |
| Egypt | 12,866 | 87,894 (+/- 12,702) ** | 6.83 |
| Uzbekistan | 640 | 17,649 (+/-  3,255) ** | 27.58 |
| Tajikistan | 90 | 8,997 (+/-  1,373) ** | 99.97 |

*As of April 21, 2021(*2*) from JHU
** Most recent values as of April 2021(*3*)
***Estimate from prior work (*42*, *62*)
****See prior work estimating excess death in China (*42*), and the World Mortality Dataset's reporting that the Chinese government is unable to provide excess death numbers: *"We are sorry to inform you that we do not have the data you requested"* (*3*).

**Table 1: Examples of severe underreporting of Covid-19 deaths in countries with lower democracy indices**
Notable examples of underreporting are shown for countries and regions with low democracy indices. Undercount ratios in especially low democracy index countries frequently approach or surpass an order of magnitude in discrepancy.

Additionally, visual inspection reveals that countries are clustered by geographic regions in the undercount versus national indices plots (Fig. 1). Geographical groups are observed in the data, with countries from the region of Eastern Europe, the Middle East, and northern Africa (and South Africa) having among the worst underreporting of the number of Covid-19 deaths. Though the majority of the African continent is unfortunately not accessible in the World Mortality Dataset at this time (*3*), it is possible that these countries may have the similar tendency of undercounting COVID-19 deaths. In contrast, Southeast and East Asian countries, Oceania, and Western democracies generally have smaller values of the undercount ratios of Covid-19 statistics. Also note the two most populous countries in the world (China and India) are missing from the World Mortality Database due to officially stated inability to provide excess death counts(*3*). Other researchers have investigated China's excess deaths separately, finding at the province-scale Hubei (the location of Wuhan, the suspected approximate geographic origin of Covid-

19) appears to have been perhaps substantially misrepresenting their Covid-19 deaths early in the pandemic (*62*), but at the national scale Chinese excess deaths are not inconsistent with reported official deaths (*42*). In contrast, India started the pandemic with lower death rates but was predicted to have explosive growth in Covid-19 death counts by early 2021 due to migrant flows, poor medical infrastructure and high population density (*63*), a prediction which unfortunately has come true during the time of writing this article (*2*). Due to the highly volatile pandemic situation in India right now, any excess death estimates would be challenging (*64*).

*Covariation of COVID-19 Deaths with Democracy Index*

Next, we tested whether national indices correlate with Covid-19 deaths (Fig. 2). Excess death per 100,000 was used instead of absolute excess death, as absolute death counts are highly correlated with population (for this data: Pearson coefficient of 0.843, p ~ $10^{-24}$). In contrast, excess death per 100,000 is uncorrelated with population (Pearson coefficient of 0.145, p = 0.181). Using this normalized excess death measure of Covid-19 deaths, we find that the EIU democracy index correlation is actually more consistent with democracies having slightly fewer excess deaths during the pandemic than authoritarian-leaning countries. In contrast, the Polity5 democracy index and state fragility index have no correlation with excess deaths (Fig. 2).

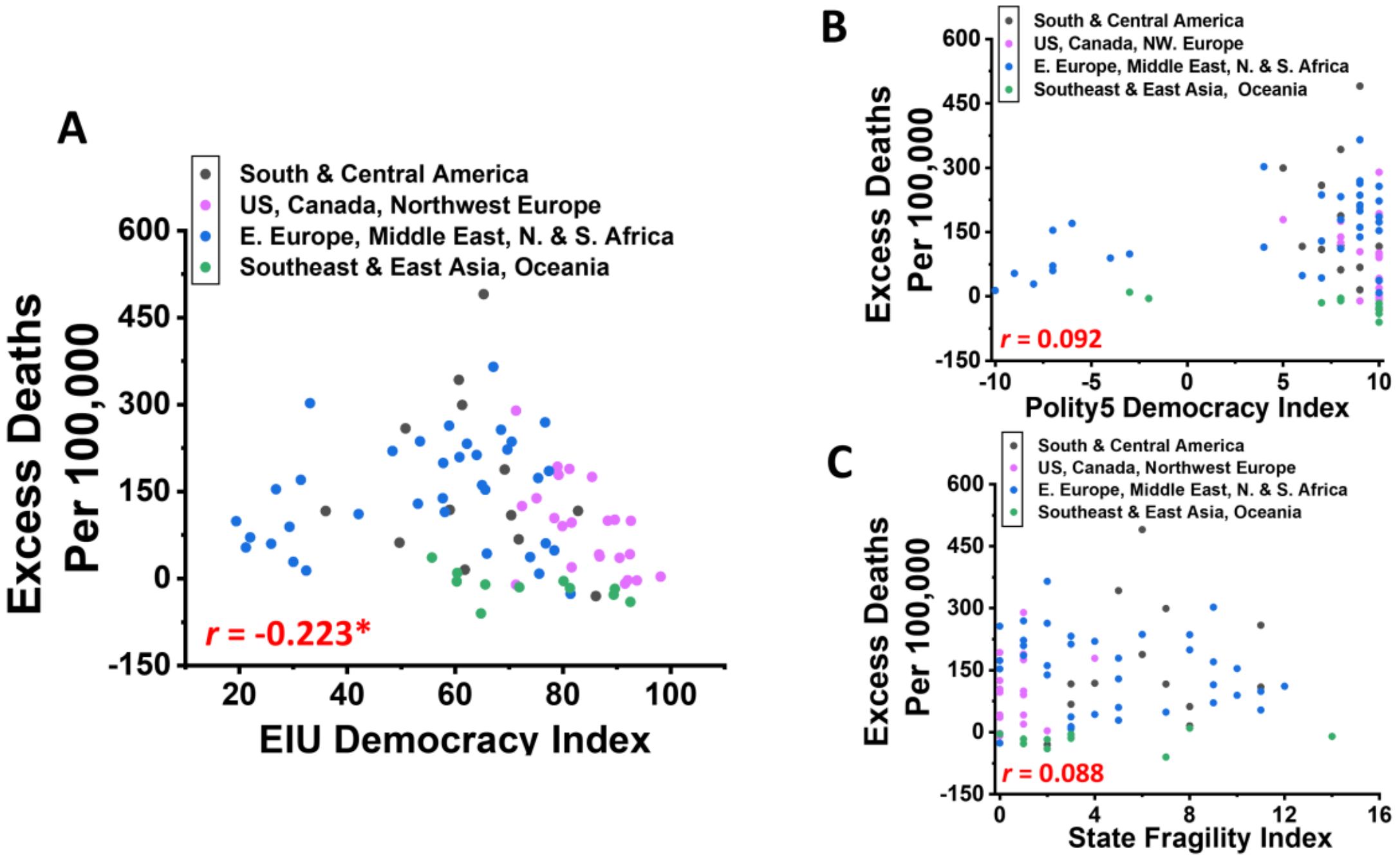

**Figure 2: Excess deaths versus national indices**

(A) The number of excess deaths per 100,000 during the pandemic versus the EIU democracy index, with a weak negative correlation observed ($p$ = 0.043) ($N$=83). (B) The number of excess deaths per 100,000 during the pandemic versus the Polity5 democracy index, with no correlation observed (*N*=80). In (A) and (B) a higher democracy index means a more democratic government, and a lower index means a more authoritarian government (*34*). (C) The number of excess deaths per 100,000 during the pandemic versus the state fragility index, with no correlation observed (*N*=81). A lower state fragility index means a more stable country.

Significant geographic clustering is seen in the excess deaths as well. Thus political systems' effects on Covid-19 deaths per country cannot be easily separated from geographic variation of the relevant parameters (e.g. mobility (*65*)) or cultural effects (*19*, *23*, *65*, *66*).

*Democracy Index and State Fragility Index*

The two democracy indices and state fragility index are all highly correlated with undercounting of Covid-19 deaths, with more fragile states and less democratic countries having worse undercounting. However, the fact that the multidimensional state fragility index correlates with the underreporting suggests that the causes of underreporting may not be purely political, assuming that democracy indices most strongly capture the political dimension (*36*). Follow-up work is necessary to explore this direction however. Additionally, weaker negative correlation with Covid-19 deaths is found only with the EIU democracy index, and no correlations are found between excess deaths and the Polity5 democracy index or state fragility index, suggesting the causes of Covid-19 deaths go well beyond the political system, as cautioned by prior work (*1*, *12*, *18*).

To examine the relationship between the democracy indices and state fragility index, we conducted additional brief analyses of correlations among these indices (Fig. 3). As expected, the two democracy indices were highly correlated as they both measured the same political dimension, but the fragile state index which incorporates four dimensions of security, economics, politics and social is less correlated with at least the Polity5 democracy index.

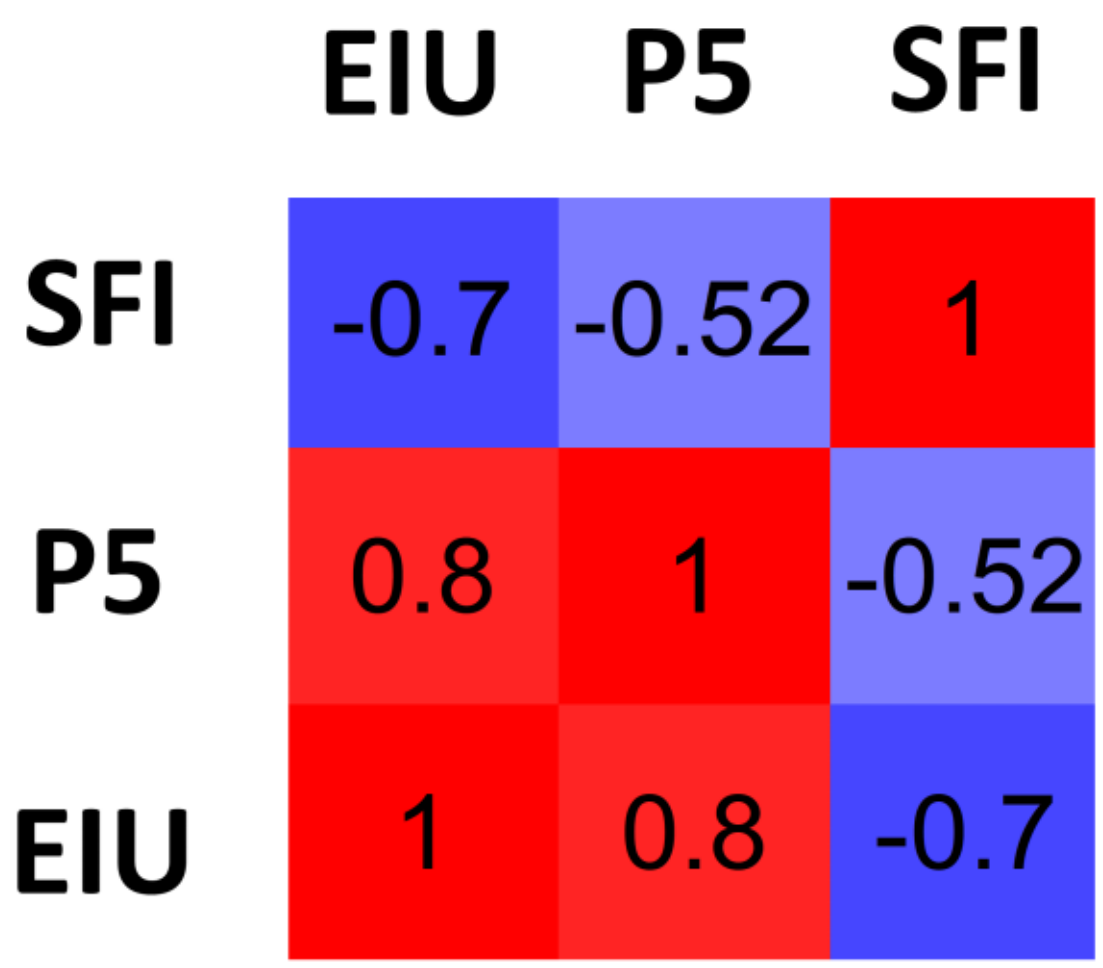

**Figure 3: Correlation matrix between national indices**
The correlation matrix between among the three national indices studied in this work: "P5" as the Center for Systemic Peace (Polity5) democracy index, "SFI" as the Center for Systemic Peace state fragility index, and "EIU" as the Economist Intelligence Unit democracy index. As expected P5 and EIU are highly positively correlated as the both measure the same political dimension. But the SFI may capture additional information beyond the political dimension, by definition, and is expected to be less correlated with a purely political democracy index.

## Conclusions

Here we have assessed whether the popular John Hopkins CSSE Database used by many researchers in the studies of the COVID-19 pandemic (*2*) can be used to analyze the causal effects of political system orientation (authoritarian versus democratic) on Covid-19 death outcomes. From the big picture perspective, as noted by prior work, democracies like New Zealand, Australia, Taiwan, Singapore, etc. performed among the best in the world in the pandemic (*7*, *67*, *68*), often with both rapid and effective Covid-19 containment and fairly popular, transparent public health strategies (*7*). Democracies also universally led the most successful vaccine development projects which are now being used to lift the world out of the pandemic crisis (*69*). More importantly, there are general suspicions that the authoritarian governments are not reporting the deaths due to the COVID-19 infections.

Our analysis here shows that skepticism of the accuracy of more fragile state and authoritarian countries' official Covid-19 death counts is justified. After comparing the official Covid-19 death counts to excess death counts, countries and regions with lower democracy indices and higher fragile state indices were found to have significantly more undercounting of Covid-19 deaths, suggesting a complicated combination of political, economic, social and security factors may be involved in inaccurate Covid-19 records. Thus, using official government records of Covid-19 deaths in global scale analyses is highly problematic from the data quality perspective.

Our results also provide some evidence that there is no obvious "Curse of Democracy" as claimed by Narita et al. (*28*) and others (*29–31*). In fact, the opposite trend seems to be present, a "Curse of Authoritarianism" (or conversely a "Blessing of Democracy", at least in terms of information veracity and possibly in Covid-19 death counts). This is not to say that all democracies performed well at all stages of the pandemic, or that all authoritarian countries performed badly at all stages (*70*). For example, the United States had one of the highest total Covid-19 death counts in the world (*2*) (albeit under an unusually authoritarian American president (*71*, *72*), and now that death count is being rivaled by India (*2*)), though the United States also now has one of the fastest vaccination rates in the world now (*73*). On the other hand, lower democracy index China contained the pandemic rapidly and effectively at the national-scale by the spring of 2020, despite beginning with highly detrimental information censorship in the early stages of the pandemic (*4*, *42*, *74*).

Generally, our analyses also suggest democracies may have performed at least slightly better at the global scale than authoritarian countries, in terms of both accurate medical record keeping and possibly in Covid-19 death mitigation. We thus urge public health researchers in this area to consider the implications of the political and governing systems on the data of the COVID-19 deaths. To address the risks of using the distorted data, we recommend employing both excess deaths and official government-reported deaths in any (especially causal) research analyses using Covid-19 deaths as an outcome variable.

**Author Contributions**

All authors analyzed the data and contributed to the manuscript.

**Funding & Acknowledgements**

R.P.B. and R.A. were funded by a research grant from Toyota Motor Corporation (LP-3009219). K.I. was funded by JSPS KAKENHI Grant Number 19K21701 and 17H04780.

**Conflicts of Interest**

R.P.B and R.A are researchers at RIKEN, which is a non-profit research institute in Japan supported by the Japanese government. However, this work is entirely the views of the authors only. R.P.B and R.A also receive funding partly from the Toyota Motor

## Ethics Declaration

All institutional, national (Japan), and international research standards were followed for this work.

## SUPPLEMENTARY INFORMATION

Below is the raw data used to make the figures. Columns are defined as follows:

*Country/Region:* The name of the country or region.

*Data Until:* The most recent date the World Mortality Dataset was updated for that country.

*Excess Deaths:* The current cumulative number of excess deaths per country since the start of 2020 from the World Mortality Dataset (*3*).

*Official Deaths:* The official (government-reported) cumulative number of Covid-19 deaths since the start of the pandemic, per country, from the John Hopkins CSSE database, (accessed April 21, 2021) (*2*).

*EIU:* The Economist Intelligence Unit democracy index (2021) (*34*).

*Polit5:* The Center for Systemic Peace democracy index (most current values), Polity5 project (the "polity" variable column in the original dataset) (*35*).

*SFI:* The Center for Systemic Peace state fragility index (most current values) (*35*).

*Undercount:* The ratio of Excess Deaths to Official Deaths. If the country has negative excess deaths, then the ratio is set to 1 (i.e. no discrepancy).

Note: Additional columns are provided in the original World Mortality Dataset and John Hopkins CSSE database, for more information on the excess death and official Covid-19 death columns respectively, if these datasets are downloaded from the source websites. The columns presented below are selected from the larger datasets as they are the variable columns used to produce the figures in this manuscript.

---------------------------- DATASET START --------------------------------------------

| Country / Regions | Data until | Excess deaths | Excess death per 100k | Official deaths | EIU | Polit5 | SFI | Under-count |
|---|---|---|---|---|---|---|---|---|
| Albania | 12/31/2020 | 6004 | 209.5 | 2358 | 60.8 | 9 | 1 | 2.54623 |
| Armenia | 2/28/2021 | 6983 | 236.6 | 3944 | 53.5 | 7 | 6 | 1.77054 |
| Australia | 12/27/2020 | -4451 | -17.8 | 910 | 89.6 | 10 | 2 | 1 |
| Austria | 4/4/2021 | 8528 | 96.5 | 9997 | 81.6 | 10 | 0 | 0.85306 |
| Azerbaijan | 12/31/2020 | 15309 | 154 | 4235 | 26.8 | -7 | 10 | 3.61488 |
| Belarus | 6/30/2020 | 5689 | 60 | 2453 | 25.9 | -7 | 5 | 2.3192 |
| Belgium | 4/4/2021 | 15840 | 138.5 | 23867 | 75.1 | 8 | 2 | 0.66368 |
| Bolivia | 2/28/2021 | 29393 | 258.9 | 12731 | 50.8 | 7 | 11 | 2.30877 |
| Bosnia and Herzegovina | 12/31/2020 | 7307 | 219.8 | 8082 | 48.4 | -- | 4 | 0.90411 |
| Brazil | 3/31/2021 | 393865 | 188 | 381475 | 69.2 | 8 | 6 | 1.03248 |
| Bulgaria | 4/11/2021 | 25643 | 365 | 15618 | 67.1 | 9 | 2 | 1.64189 |
| Canada | 1/3/2021 | 15520 | 41.9 | 23761 | 92.4 | 10 | 0 | 0.65317 |
| Chile | 4/4/2021 | 21837 | 116.6 | 25353 | 82.8 | 10 | 3 | 0.86132 |
| Colombia | 1/17/2021 | 54286 | 109.3 | 69596 | 70.4 | 7 | 11 | 0.78002 |
| Costa Rica | 12/31/2020 | 965 | 19.3 | 3115 | 81.6 | 10 | 1 | 0.30979 |
| Croatia | 2/28/2021 | 6581 | 161 | 6692 | 65 | 9 | 2 | 0.98341 |
| Cyprus | 3/28/2021 | 98 | 8.3 | 295 | 75.6 | 10 | 3 | 0.3322 |
| Czechia | 3/14/2021 | 28630 | 269.3 | 28711 | 76.7 | 9 | 1 | 0.99718 |
| Denmark | 4/11/2021 | -508 | -8.8 | 2466 | 91.5 | 10 | 0 | 1 |
| Ecuador | 4/11/2021 | 51121 | 299.2 | 17804 | 61.3 | 5 | 7 | 2.87132 |
| Egypt | 11/30/2020 | 87894 | 89.3 | 12866 | 29.3 | -4 | 10 | 6.83149 |
| El Salvador | 8/31/2020 | 7596 | 118.3 | 2086 | 59 | 8 | 4 | 3.64142 |
| Estonia | 4/11/2021 | 1376 | 104.1 | 1109 | 78.4 | 9 | 0 | 1.24076 |
| Finland | 4/4/2021 | -159 | -2.9 | 899 | 92 | 10 | 0 | 1 |
| France | 4/4/2021 | 60385 | 90.2 | 102046 | 79.9 | 10 | 1 | 0.59174 |
| Georgia | 12/31/2020 | 4804 | 128.9 | 3971 | 53.1 | 7 | 5 | 1.20977 |
| Germany | 4/4/2021 | 34318 | 41.4 | 80938 | 86.7 | 10 | 1 | 0.424 |
| Greece | 2/28/2021 | 3966 | 37 | 9713 | 73.9 | 10 | 3 | 0.40832 |
| Greenland | 12/31/2020 | -16 | -27.8 | -- | -- | -- | -- | -- |
| Guatemala | 12/27/2020 | 10681 | 61.9 | 7309 | 49.7 | 8 | 8 | 1.46135 |
| Hong Kong | 2/28/2021 | 2691 | 36.1 | -- | 55.7 | -- | -- | -- |
| Hungary | 3/21/2021 | 14980 | 153.2 | 25787 | 65.6 | 10 | 0 | 0.58091 |
| Iceland | 3/21/2021 | -12 | -3.3 | 29 | 93.7 | -- | -- | 1 |
| Iran | 9/21/2020 | 58092 | 71 | 67913 | 22.0 | -7 | 9 | 0.85539 |
| Ireland | 2/28/2021 | 1743 | 35.8 | 4856 | 90.5 | 10 | 0 | 0.35894 |
| Israel | 3/21/2021 | 4296 | 48.4 | 6346 | 78.4 | 6 | 7 | 0.67696 |
| Italy | 1/31/2021 | 112047 | 185.4 | 117997 | 77.4 | 10 | 1 | 0.94957 |
| Jamaica | 11/30/2020 | -315 | -10.7 | 744 | 71.3 | 9 | 3 | 1 |
| Japan | 2/28/2021 | -20842 | -16.5 | 9737 | 81.3 | 10 | 1 | 1 |
| Kazakhstan | 2/28/2021 | 31095 | 170.2 | 3301 | 31.4 | -6 | 9 | 9.41987 |
| Kosovo | 2/28/2021 | 3304 | 179 | 2105 | -- | 8 | 5 | 1.5696 |
| Kyrgyzstan | 1/31/2021 | 7028 | 111.2 | 1561 | 42.1 | 8 | 12 | 4.50224 |
| Latvia | 4/4/2021 | 2409 | 125 | 2079 | 72.4 | 8 | 0 | 1.15873 |
| Liechtenstein | 2/28/2021 | 50 | 131.1 | 56 | -- | -- | -- | 0.89286 |
| Lithuania | 4/4/2021 | 8110 | 289.5 | 3802 | 71.3 | 10 | 1 | 2.13309 |
| Luxembourg | 3/14/2021 | 231 | 38 | 788 | 86.8 | 10 | 0 | 0.29315 |
| Macao | 2/28/2021 | -17 | -2.8 | -- | -- | -- | -- | -- |
| Malaysia | 12/31/2020 | -4728 | -15 | 1400 | 71.9 | 7 | 3 | 1 |
| Malta | 3/21/2021 | 294 | 60.6 | 411 | 76.8 | -- | -- | 0.71533 |
| Mauritius | 12/31/2020 | -333 | -26.3 | 15 | 81.4 | 10 | 0 | 1 |
| Mexico | 3/7/2021 | 432104 | 342.4 | 213597 | 60.7 | 8 | 5 | 2.02299 |
| Moldova | 12/31/2020 | 5393 | 199.3 | 5643 | 57.8 | 9 | 8 | 0.9557 |
| Mongolia | 3/31/2021 | -1905 | -60.1 | 56 | 64.8 | 10 | 7 | 1 |
| Montenegro | 1/31/2021 | 861 | 138.4 | 1444 | 57.7 | 9 | 2 | 0.59626 |
| Netherlands | 4/11/2021 | 17481 | 101.4 | 17202 | 89.6 | 10 | 0 | 1.01622 |
| New Zealand | 3/28/2021 | -1951 | -40.3 | 26 | 92.5 | 10 | 2 | 1 |
| Nicaragua | 8/31/2020 | 7528 | 116.4 | 181 | 36 | 6 | 7 | 41.59116 |
| North Macedonia | 1/31/2021 | 5490 | 263.6 | 4556 | 58.9 | 9 | 2 | 1.205 |
| Norway | 1/3/2021 | 170 | 3.2 | 734 | 98.1 | 10 | 2 | 0.23161 |
| Oman | 3/31/2021 | 1388 | 28.7 | 1926 | 30 | -8 | 5 | 0.72066 |
| Panama | 12/31/2020 | 2824 | 67.6 | 6196 | 71.8 | 9 | 3 | 0.45578 |
| Paraguay | 12/31/2020 | 1056 | 15.2 | 5561 | 61.8 | 9 | 8 | 0.18989 |
| Peru | 4/18/2021 | 156839 | 490.3 | 57954 | 65.3 | 9 | 6 | 2.70627 |
| Philippines | 11/30/2020 | -10986 | -10.3 | 16265 | 65.6 | 8 | 14 | 1 |
| Poland | 4/11/2021 | 97446 | 256.6 | 63473 | 68.5 | 10 | 0 | 1.53524 |
| Portugal | 4/4/2021 | 19827 | 192.8 | 16952 | 79 | 10 | 0 | 1.1696 |
| Qatar | 2/28/2021 | 377 | 13.6 | 400 | 32.4 | -10 | 3 | 0.9425 |
| Romania | 2/28/2021 | 41478 | 213.1 | 26793 | 64 | 9 | 3 | 1.54809 |
| Russia | 2/28/2021 | 443695 | 302.4 | 104937 | 33.1 | 4 | 9 | 4.2282 |
| San Marino | 2/28/2021 | 89 | 263.1 | 88 | -- | -- | -- | 1.01136 |
| Serbia | 2/28/2021 | 16222 | 232.3 | 6095 | 62.2 | 8 | 3 | 2.66153 |
| Singapore | 12/31/2020 | -289 | -5.1 | 30 | 60.3 | -2 | 3 | 1 |
| Slovakia | 2/28/2021 | 12110 | 222.3 | 11304 | 69.7 | 10 | 1 | 1.0713 |
| Slovenia | 3/21/2021 | 3594 | 173.3 | 4176 | 75.4 | 10 | 0 | 0.86063 |
| South Africa | 4/11/2021 | 136380 | 236 | 53940 | 70.5 | 9 | 8 | 2.52836 |
| South Korea | 1/31/2021 | -2251 | -4.4 | 1808 | 80.1 | 8 | 0 | 1 |
| Spain | 3/28/2021 | 88457 | 189 | 77364 | 81.2 | 10 | 1 | 1.14339 |
| Sweden | 2/28/2021 | 10117 | 99.4 | 13863 | 92.6 | 10 | 0 | 0.72978 |
| Switzerland | 4/4/2021 | 8484 | 99.7 | 10546 | 88.3 | 10 | 1 | 0.80448 |
| Taiwan | 3/31/2021 | -6581 | -27.9 | 11 | 89.4 | 10 | 1 | 1 |
| Tajikistan | 12/31/2020 | 8997 | 98.9 | 90 | 19.4 | -3 | 11 | 99.96667 |
| Thailand | 3/31/2021 | 6655 | 9.6 | 110 | 60.4 | -3 | 8 | 60.5 |
| Transnistria | 2/28/2021 | 899 | 192.3 | -- | -- | -- | -- | -- |
| Tunisia | 2/14/2021 | 4977 | 43 | 9993 | 65.9 | 7 | 4 | 0.49805 |
| Ukraine | 2/28/2021 | 47942 | 114.8 | 42565 | 58.1 | 4 | 9 | 1.12632 |
| United Kingdom | 4/4/2021 | 116490 | 175.3 | 127577 | 85.4 | 8 | 1 | 0.9131 |
| United States | 2/21/2021 | 584332 | 178.9 | 569402 | 79.2 | 5 | 4 | 1.02622 |
| Uruguay | 7/26/2020 | -1044 | -30.3 | 2083 | 86.1 | 10 | 2 | 1 |
| Uzbekistan | 12/31/2020 | 17649 | 53.6 | 640 | 21.2 | -9 | 11 | 27.57656 |

---------------------------- DATASET END --------------------------------------------